\documentclass[preprint,aps]{revtex4}

\usepackage{graphicx}

\begin{document}

\title{Anomalous High-Energy Waterfall-Like Electronic Structure in 5{\it d} Transition Metal Oxide Sr$_2$IrO$_4$ with a Strong Spin-Orbit Coupling}

\author{Yan Liu$^{1}$, Li Yu$^{1,*}$, Xiaowen Jia$^{1}$, Jianzhou Zhao$^{1}$, Hongming Weng$^{1}$, Yingying Peng$^{1}$, Chaoyu Chen$^{1}$, Zhuojin Xie$^{1}$, Daixiang Mou$^{1}$, Junfeng He$^{1}$, Xu Liu$^{1}$, Ya Feng$^{1}$, Hemian Yi$^{1}$, Lin Zhao$^{1}$, Guodong  Liu$^{1}$, Shaolong He$^{1}$, Xiaoli Dong$^{1}$, Jun Zhang$^{1}$, Zuyan Xu$^{3}$, Chuangtian Chen$^{3}$, Gang Cao$^{4}$, Xi Dai$^{1}$, Zhong Fang$^{1}$ and X. J. Zhou$^{1,2,*}$
}

\affiliation{
\\$^{1}$Beijing National Laboratory for Condensed Matter Physics, Institute of Physics,
Chinese Academy of Sciences, Beijing 100190, China
\\$^{2}$Collaborative Innovation Center of Quantum Matter, Beijing, China
\\$^{3}$Technical Institute of Physics and Chemistry, Chinese Academy of Sciences, Beijing 100190, China
\\$^{4}$Department of Physics and Astronomy, University of Kentucky, Lexington, KY 40506
}
\date{January 20, 2015}




\maketitle

\newpage

{\bf The layered {\it 5d} transition metal oxides like Sr$_2$IrO$_4$ have attracted significant interest recently due to a number of exotic and new phenomena induced by the interplay between the spin-orbit coupling, bandwidth W and on-site Coulomb correlation U. In contrast to a metallic behavior expected from the Mott-Hubbard model due to more spatially extended {\it 5d} orbitals and moderate U, an insulating ground state has been observed in Sr$_2$IrO$_4$. Such an insulating behavior can be understood by an effective J$_{eff}$=1/2 Mott insulator model by incorporating both electron correlation and strong spin-orbital coupling, although its validity remains under debate at present. In particular, Sr$_2$IrO$_4$ exhibits a number of similarities to the high temperature cuprate superconductors in the crystal structure, electronic structure, magnetic structure, and even possible high temperature superconductivity that is predicted in doped Sr$_2$IrO$_4$.  Here we report a new observation of the anomalous high energy electronic structure in Sr$_2$IrO$_4$. By taking high-resolution angle-resolved photoemission measurements on Sr$_2$IrO$_4$ over a wide energy range, we have revealed for the first time that the high energy electronic structures show unusual nearly-vertical bands that extend over a large energy range. Such anomalous high energy behaviors resemble the high energy waterfall features observed in the cuprate superconductors, adding one more important similarity between these two systems. While strong electron correlation plays an important role in producing high energy waterfall features in the cuprate superconductors, the revelation of the high energy anomalies in Sr$_2$IrO$_4$ points to a novel route in generating exotic electronic excitations from the strong spin-orbit coupling and a moderate electron correlation.
}

The transition metal oxides exhibit rich exotic physical properties such as high temperature superconductivity and colossal magnetoresistance that have become a central theme of modern condensed matter physics\cite{MImada,PLeeRMP}. The insulating ground state of the {\it 3d} transition metal oxides can generally be understood by the strong on-site Coulomb repulsion U, relative to its bandwidth W (U$\gg$W), as proposed in the Mott-Hubbard model\cite{MImada}. An insulator-metal transition can occur when W$\geq$U. In comparison, in the {\it 5d} transition metal oxides, the electron correlation is expected to become less strong due to the more spatially extended {\it 5d} orbitals and a metallic ground state is expected\cite{LFMattheiss1969,LFMattheiss1976}. It is thus surprising when it was found that the prototypical {\it 5d} compound Sr$_2$IrO$_4$ is an antiferromagnetic insulator below the Neel temperature T$_N$$\sim$240 K\cite{Crawford,GCao1998,BJKim2008PRL,SFujiyama}. One popular scenario for the novel insulating ground state of Sr$_2$IrO$_4$ is the J$_{eff}$=1/2 Mott insulator model driven by spin-orbit coupling\cite{BJKim2008PRL}. In this model, five {\it 5d} electrons occupy the t$_{2g}$ orbitals which are split into a fully-filled J$_{eff}$=3/2 quartet band with lower energy and a half-filled  doublet band with J$_{eff}$=1/2 close to the Fermi level (E$_F$) by strong spin-orbit coupling.  Assuming that the width of the J$_{eff}$=1/2 band is narrow, a moderate Coulomb repulsion U can open up a gap, giving rise to the so-called J$_{eff}$=1/2 Mott insulating ground state\cite{BJKim2008PRL}. While a couple of experimental results are consistent with this scenario\cite{BJKim2008PRL,SJMoon2008,BJKim2009,KIshii,BMWojek,QWang,SMoser,JNichols,JXDai,YCao2014}, this picture is challenged due to the fact that the electron hopping energy ($\sim$0.3 eV) is not negligible which suggests the J$_{eff}$=1/2 state alone cannot fully describe the system\cite{HWatanabe}.  Moreover, on the formation mechanism of the insulating ground state in Sr$_2$IrO$_4$, an alternative magnetic or Slater-type origin, instead of the Mott-type\cite{BJKim2008PRL}, has been proposed\cite{RArita} with some experimental support\cite{DHsieh,AYamasaki,QLi}.   These leave the very nature of the insulating ground state in Sr$_2$IrO$_4$ open and it is important to study and understand the electronic structure of Sr$_2$IrO$_4$.

Angle-resolved  photoemission spectroscopy (ARPES) is a powerful tool to directly probe the low energy electronic structures of solid materials\cite{Damascelli}. The existing ARPES results on Sr$_2$IrO$_4$ and related compounds within a relatively narrow energy window appear to agree with the J$_{eff}$=1/2 model\cite{BJKim2008PRL,QWang,BMWojek,SMoser,YCao2014}.  In this paper, we report the observation of unusual high energy bands in Sr$_2$IrO$_4$.  Our comprehensive angle-resolved photoemission study over a wide energy window reveals for the first time nearly-vertical bands in Sr$_2$IrO$_4$. Such  exotic bands can not be understood in terms of the band structure calculations; they cannot be understood within the simple J$_{eff}$=1/2 Mott insulator model or any other existing theoretical models either. The observed high energy anomaly resemble the unusual high energy waterfall bands discovered in the high temperature cuprate superconductors\cite{RonningPRB,GrafPRL,XiePRL,VallaPRL,NonHighEKink,ChangPRB,InosovPRL,ZhangPRL,MoritzNJP,IkedaPRB}. These observations point to the significant role of the strong spin-orbit coupling, together with a moderate electron correlation, in giving rise to new high energy excitations in the {\it 5d} transition metal oxides.

The Sr$_2$IrO$_4$ single crystals were synthesized by  flux method\cite{GCao1998}. High-resolution angle-resolved photoemission measurements were carried out on our lab system equipped with a Scienta R4000 electron energy analyzer\cite{GDLiu}. We use helium discharge lamp as the light source that can provide photon energy of h$\upsilon$=21.218eV (helium I). The overall energy resolution was set at 20 meV. The angular resolution is  $\sim$0.3 degree. The Fermi level is referenced by measuring on a clean polycrystalline gold that is electrically connected to the sample. The sample was cleaved {\it in situ} and measured at $\sim$20 K in ultra-high vacuum with a base pressure better than 5$\times$10$^{-11}$ Torr. The measurements were carried out on different samples for several times and the results are reproducible.

Figure 1 shows the constant energy contours of Sr$_2$IrO$_4$ at different binding energies. No spectral weight is present at the Fermi level (not shown in Fig. 1), consistent with the insulating nature of Sr$_2$IrO$_4$\cite{Crawford,GCao1998,BJKim2008PRL}. At a binding energy of 0.2 eV, the spectral weight appears first as a circular spot around the X($\pi$,0) and its equivalent locations (Fig. 1a).  Further increase of the binding energy to 0.4 eV results in the enlargement of the spot into a square-shape and the emergence of spectral weight near the $\Gamma$ (0,0) point (Fig. 1b). When the binding energy increases to 0.8 eV, the strong spectral weight near X points vanishes with a formation of a few disconnected patches around X, while the spectral weight near $\Gamma$ exhibits a petal-like shape with four leaves (Fig. 1c). The measured constant energy contours at low binding energy (0$\sim$0.4 eV) are consistent with those reported before\cite{BJKim2008PRL,QWang}.  The constant energy contours at an intermediate  binding energy (e.g., 0.4 eV)  are also consistent with the band structure calculations (Fig. 1d) that include both the on-site Coulomb repulsion U and the spin-orbit coupling\cite{BJKim2008PRL}.  In terms of the spin-orbit-coupling-driven Mott insulator model\cite{BJKim2008PRL}, the unoccupied states are mainly the J$_{eff}$=1/2 state, while the occupied states are a mixture of the J$_{eff}$=1/2 and 3/2 states.  Due to the strong spin-orbit coupling, the topmost low energy valence state at X is more with J$_{eff}$=1/2 character ($\beta$ sheet near X in Fig. 1d), while the topmost low energy valence state at $\Gamma$ is more with J$_{eff}$=3/2 character ($\alpha$ sheet near $\Gamma$ in Fig. 1d)\cite{BJKim2008PRL}.  The consistency of the low energy electronic structure with the previous reports and the band structure calculations lays a foundation for our following investigation of high binding energy electronic structure in Sr$_2$IrO$_4$.

At high binding energies, we find that the electronic structure of Sr$_2$IrO$_4$ is quite unusual. Figure 2 shows band structure along two high-symmetry momentum cuts covering a large energy range till $\sim$6 eV: one cut is across $\Gamma$ (Fig. 2a-d), the other is across X (Fig. 2e-h). Here we show both the original data (Fig. 2a and 2e), and their corresponding momentum-(Fig. 2b and 2f) and energy-second-derivative (Fig. 2c and 2g) images. The second-derivative images help to highlight the band structure more clearly although many features are already clear in the original data. Since momentum-second derivative image may miss the flat horizontal bands while the energy-second derivative image may miss the vertical bands, the energy- and momentum-second-derivative images are complementary to each other to provide a full picture. As seen in Fig. 2, at low binding energy (0$\sim$1 eV),  two prominent bands are observed labeled as $\alpha_0$ and $\beta_0$ (Fig. 2c and 2g) that are consistent with the previous reports\cite{BJKim2008PRL,QWang}. However, at higher binding energy, the electronic structure becomes quite unusual. First, the momentum-second derivative images and energy-second-derivative images give rather different band structures for both the $\Gamma$ and X momentum cuts. Second, as seen in Fig. 2b, a clear vase-shaped band (labeled as $\alpha_1$ in Fig. 2b) and a vertical waterfall-like band (labeled as $\alpha_2$ in Fig. 2b) are observed around the $\Gamma$ point. The vertical band structure is present even beyond 3 eV up to $\sim$6 eV (Fig. 2b). Note that these features are not due to the artifact of the momentum second-derivative image because they are already clear in the original data (Fig. 2a). Such features can also be identified clearly in the momentum distribution curves (MDCs) where the peaks corresponding to $\alpha_1$ and $\alpha_2$ bands are marked (Fig. 2d).  Similar behaviors are observed for the momentum cut across the X point (Fig. 2e) where nearly vertical band structures (labeled as $\alpha_3$ and $\beta_1$ in Fig. 2f) are observed up to 4 eV, and another set of vertical bands are seen even up to 6.5 eV (Fig. 2f).

The unusual high energy electronic structure of Sr$_2$IrO$_4$ is present over a large momentum space.  Figure 3 shows the detailed momentum evolution of the high energy electronic structure: one is near the $\Gamma$ region (Fig. 3a-d) and the other near the X(0,$\pi$) region (Fig. 3e-h). While the energy-second-derivative images (Fig. 3d and 3h) show normal two bands ($\alpha_0$ and $\beta_0$) in the covered energy range as already seen in Fig. 2, nearly vertical bands are observed in the momentum-second-derivative images (Fig. 3c and 3g) in both cases for different momentum cuts. Furthermore, the constant energy contours exhibit dramatic evolution with the binding energy (Fig. 3a and 3e).  The spectral weight distribution around the $\Gamma$ point (Fig. 3a) changes from a pocket centered at $\Gamma$ at a binding energy of 0.4 eV, to butterfly-shaped at 0.6 eV and 0.8 eV, to big-X-shaped at 1.2 eV and to dumbbell-shaped at 2.0 eV and 2.4 eV. It is interesting to note that the spectral weight distribution shows discrete four strong spots at 0.6 eV and 0.8 eV, other than a continuous contour. From Fig. 3c and 3d, it becomes clear that the drastic spectral distribution change with the binding energy above 1.0 eV is directly related with the presence of the nearly-vertical $\alpha_1$ and $\alpha_2$ bands. It is also clear from Fig. 3b that, moving away from the cut across $\Gamma$ (cut 1), the vase-shaped band and vertical structure persist for the cuts 2 and 3.  The same is true for the X point constant energy contours (Fig. 3e) and the momentum-dependent band structures (Fig. 3f-h). First, the constant energy contours near X also exhibit an obvious evolution with the binding energy (Fig. 3e). Second, the vertical bands are present over a large area of momentum space near X (Fig. 3g).

Figure 4 summarizes the band structure of Sr$_2$IrO$_4$ measured along three typical high-symmetry momentum cuts (Fig. 4c-e). For comparison, the band structures of Sr$_2$IrO$_4$ in the antiferromagnetic state are also calculated using the DMFT method with U=2.5, J=0 and $\beta$=100 (Fig. 4a and 4b). In the calculated band structure (Fig. 4a), the electronic states between the Fermi level and 3 eV binding energy are mainly from the Iridium's t$_{2g}$ orbitals (white lines in Fig. 4a) while above 3 eV binding energy the contribution is mainly from the oxygen p orbitals (yellow lines in Fig. 4a).  In addition, the orbital-resolved density of states (DOS) is also calculated where the peak position of $\alpha$$_0$ and $\beta$$_0$ bands are well resolved (Fig. 4b).  In the measured band structure, from the energy-second-derivative image (Fig. 4d),  two bands are clearly observed that are marked as $\alpha_0$ and $\beta_0$ between E$_F$ and $\sim$1 eV binding energy. These two bands show good agreement with the band structure calculations (Fig. 4a) and previous reports\cite{BJKim2008PRL,QWang}. Also above 3 eV binding energy, the observed bands in the energy-second-derivative image (Fig. 4d) can find some good correspondence in the calculated band structure (Fig. 4a). The most dramatic difference between the measurements and calculations lies in the binding energy region above 1 eV.  As seen in Fig. 4a, a couple of energy bands from Iridium are expected from the band structure calculations within the energy range of 1$\sim$3 eV  but are not observed in the measured data (Fig. 4d). Instead, a number of nearly-vertical band features ($\alpha_1$, $\alpha_2$, $\alpha_3$ and $\beta_1$ bands in Fig. 4e) appear within this energy range that are completely absent in the calculated band structure (Fig. 4a). The same is for the 3$\sim$6 eV binding energy range where some vertical bands are observed (Fig. 4e) but are not present in the calculated band structure at all (Fig. 4a).

Further inspection of the measured band structure indicates that the new nearly-vertical high energy bands appear to have a close connection with the lower energy $\alpha_0$ and $\beta_0$ bands, as shown in Fig. 4c which summarizes all the observed bands on top of the original measured data. One can see that the three vertical bands $\alpha_1$, $\alpha_2$ and $\alpha_3$ merge into the $\alpha_0$ band at lower binding energy  while the other vertical band $\beta_1$ also merges into the lower binding energy $\beta_0$ band. The low-energy electronic structure of Sr$_2$IrO$_4$ are mainly composed of t$_{2g}$ bands that are split into two branches with the effective J$_{eff}$=1/2 and J$_{eff}$=3/2 because of the strong spin-orbit coupling\cite{BJKim2008PRL}.  It has been shown that the $\beta_0$ band is predominantly with the J$_{eff}$=1/2 character while the $\alpha_0$ band is mainly with the J$_{eff}$=3/2 character (Fig. 4a and 4b)\cite{BJKim2008PRL}.  It is interesting to note that one vertical band ($\beta_1$) emerges from the J$_{eff}$=1/2 $\beta_0$ band while three vertical bands ($\alpha_1$, $\alpha_2$ and $\alpha_3$) emerge from the J$_{eff}$=3/2 $\alpha_0$ band (Fig. 4c), consistent with the orbital degeneracy of both the J$_{eff}$ =1/2 and 3/2 bands.  These observations indicate the multi-orbital nature of the low energy electronic states in Sr$_2$IrO$_4$.  In both the J$_{eff}$ =1/2 and 3/2 bands, we have observed such waterfall-like features splitting out of the original bands,  this means that the high energy anomaly in Sr$_2$IrO$_4$ is a general feature appearing for all orbitals over a rather high energy scale.

To the best of our knowledge, such unusual high energy waterfall-like electronic structures are observed for the first time in Sr$_2$IrO$_4$. The appearance of nearly-vertical bands is quite unusual because it implies nearly infinite electron velocity if interpreted literally in the conventional band structure picture.  This is reminiscent to the high energy waterfall feature observed in the high temperature cuprate superconductors\cite{RonningPRB,GrafPRL,XiePRL,VallaPRL,NonHighEKink,ChangPRB,InosovPRL,ZhangPRL,MoritzNJP,IkedaPRB}. The high energy behaviors are similar between the cuprates and Sr$_2$IrO$_4$ in a couple of aspects.  First, the energy- and momentum-second-derivative images give different band structure\cite{GrafPRL,ZhangPRL}.  For a conventional metal, the energy- and momentum-second-derivative images are supposed to produce similar band structure. The dichotomy between them already points to an exotic behavior and the effect of strong correlation. Second, nearly vertical bands are observed in the momentum-second-derivative images. The behavior in Sr$_2$IrO$_4$ is even more dramatic in that several bands show such waterfall-like high energy features (Fig. 4e) while only one band in cuprates shows such a behavior\cite{RonningPRB,GrafPRL,XiePRL,VallaPRL,NonHighEKink,ChangPRB,InosovPRL,ZhangPRL,MoritzNJP,IkedaPRB}. Moreover, the high energy features in Sr$_2$IrO$_4$ extend over a much larger energy range (1$\sim$3 eV for Ir states) (Fig. 4) while it is in the scale of 0.4$\sim$1 eV in cuprates\cite{RonningPRB,GrafPRL,XiePRL,VallaPRL,NonHighEKink,ChangPRB,InosovPRL,ZhangPRL,MoritzNJP,IkedaPRB}. The high energy behavior in Sr$_2$IrO$_4$ is even more complicated, such as the observation of a vase-like shape near the $\Gamma$ point ( $\alpha_1$ band Fig. 2a and 2b).

The revelation of the high energy waterfall-like bands in Sr$_2$IrO$_4$ provides another system that can be used to compare and contrast with the cuprates in order to understand the origin of the high energy anomaly. In the cuprate superconductors, the high energy anomalous band has attracted extensive experimental\cite{RonningPRB,GrafPRL,XiePRL,VallaPRL,NonHighEKink,ChangPRB,InosovPRL,ZhangPRL,MoritzNJP,IkedaPRB} and theoretical interest\cite{Byczuk,Manousakis,Leigh,RSMPlasmon,RSMParamagnon,TZhou,Macridin,LJZhu,FTan,MMZemljic,PSrivastava,Alexandrov,CWeber,SBasak,SSakai,DKatagiri,BDPiazza,GMazza} although there has been no consensus reached on its origin. The prime candidate for the anomalous high energy behavior can be simply an intrinsic property of a strong electron correlation system or Mott physics\cite{NonHighEKink,Byczuk,Leigh,FTan,MMZemljic,PSrivastava,CWeber,SBasak,DKatagiri,BDPiazza}. The second possibility is due to quasiparticle scattering with some electronic or bosonic excitations, such as phonons\cite{XiePRL}, plasmons\cite{RSMPlasmon}, paramagnons\cite{VallaPRL,Macridin, RSMParamagnon,SBasak}, and other spin and charge excitaions\cite{GMazza}.  It can also be due to other novel effects such as the spin-charge separation\cite{GrafPRL}, spin polarons\cite{Manousakis}, photoemission matrix element effect\cite{InosovPRL}, charge modulations\cite{TZhou}, quantum critical fluctuation\cite{LJZhu}, in-gap band-tails\cite{Alexandrov} and so on.     Compared with the cuprates where there is a strong electron correlation\cite{PLeeRMP}, the electron correlation in {\it 5d} transition metal oxide Sr$_2$IrO$_4$ is much weaker owing to the much extended {\it 5d} orbitals\cite{BJKim2008PRL}. On the other hand, due to heavier atomic mass, the spin-orbit coupling becomes an order of magnitude stronger($\sim$0.5 eV)\cite{MMontalti} in the {\it 5d} transition metal oxides than that in their {\it 3d} counterparts ($\sim$20 meV), reaching a comparable energy scale with the on-site Coulomb repulsion U and the bandwidth W\cite{BJKim2008PRL,MGe}. This indicates that the spin-orbit coupling provides a novel tuning parameter in dictating the ground state and physical properties of the {\it 5d} transition metal oxides.  While the strong electron correlation plays an important role in producing high energy anomaly in the cuprate superconductors, the observation of the high energy anomaly in Sr$_2$IrO$_4$ provides a new scenario where the high energy anomaly can be observed in a system with a moderate or weak  electron correlation and strong spin-orbit coupling.

One further question comes to whether the moderate electron correlation or the strong spin-orbit coupling alone can produce such a high energy anomaly in Sr$_2$IrO$_4$ or it is a combined effect.  It would be surprising if a moderate electron correlation alone in Sr$_2$IrO$_4$ can cause the high energy anomaly over much larger energy scale than that in cuprates which has much stronger electron correlation although the possibility cannot be fully ruled out. There is no observation of high energy anomaly reported in systems with dominant spin-orbit coupling like simple metal Bi\cite{BiReview} or topological insulators\cite{TIReview1,TIReview2}. The anomalous high energy features can be most likely a combined effect of both the electron correlation and the spin-orbit coupling. This is consistent with the proposition that, in order to understand the insulating behavior of Sr$_2$IrO$_4$, both the on-site Coulomb interaction and strong spin-orbit coupling are necessary\cite{BJKim2008PRL}.  It is also consistent with the recent observation of a high energy anomaly in UCoGa$_5$ that exhibits a moderate electron correlation and strong spin-orbital coupling\cite{TDasHFHighE}.  Exotic quasiparticles like a composite particle has been reported lately in Sr$_2$IrO$_4$\cite{JHKim}. How the combination of the moderate electron correlation and the strong spin-orbit coupling can lead to such anomalous high energy excitations in Sr$_2$IrO$_4$ needs further theoretical and experimental investigations.

Interestingly, Sr$_2$IrO$_4$ exhibits a number of features that are similar to those of the high temperature cuprate superconductors. First, its crystal structure\cite{Crawford} is similar to that of a parent compound La$_2$CuO$_4$\cite{PLeeRMP} with a slight distortion.  Second, its insulating nature might be described by a J$_{eff}$=1/2 Mott insulator model\cite{BJKim2008PRL} that is similar to the Mott insulator model for the parent compounds of the cuprate superconductors\cite{PLeeRMP}. Third, the electron-doped Sr$_2$IrO$_4$ shows a single hole-like Fermi surface and Fermi arc\cite{YKKim2014} that are quite  reminiscent  to that found in doped cuprate superconductors\cite{Damascelli}.  It is suggested that the half-filled doublet J$_{eff}$=1/2 band would be mainly responsible for the low energy insulating physics in Sr$_2$IrO$_4$  as the role of the half-filled d$_{x2-y2}$ band in cuprate parent compounds.  The structural, electronic and magnetic similarities between Sr$_2$IrO$_4$ and the cuprate parent compound La$_2$CuO$_4$\cite{FWang} imply potential realization of superconductivity in doped Sr$_2$IrO$_4$\cite{HWatanabe2013}.  Our present observation of anomalous high energy waterfall-like feature in Sr$_2$IrO$_4$ adds one more prominent similarity to that in the cuprate superconductors.

In summary, our ARPES measurements over a wide energy window  have revealed for the first time a new phenomenon of the high energy anomalous bands in Sr$_2$IrO$_4$. It resembles the high energy waterfall feature observed in high temperature cuprate superconductors. While the low energy electron excitations in Sr$_2$IrO$_4$ can be described properly by considering both the on-site Coulomb repulsion and the strong spin-orbit coupling\cite{BJKim2008PRL},  the high energy anomalous bands cannot be understood in the framework of the existing band structure calculations. Different from the cuprate superconductors where strong electron correlation plays an important role in producing high energy anomalies, the present results in Sr$_2$IrO$_4$ provides a new scenario that high energy anomaly can occur in a system with moderate or weak electron correlation and strong spin-orbit coupling. We hope these experimental observations can stimulate further theoretical work in understanding the anomalous electronic behaviors in Sr$_2$IrO$_4$ in particular, and the high energy anomaly in other materials in general.\\


$^{*}$Corresponding authors: XJZhou@aphy.iphy.ac.cn, li.yu@iphy.ac.cn

\vspace{3mm}

\noindent {\bf Acknowledgement} XJZ thanks financial support from the NSFC (91021006 and 11334010),  the MOST of China (973 program No: 2011CB921703, 2011CBA00110, 2012CB821402, 2013CB921700 and 2013CB921904), and the Strategic Priority Research Program (B) of the Chinese Academy of Sciences (Grant No. XDB07020300).

\vspace{3mm}

\noindent {\bf Author Contributions}\\
X.J.Z. and Y.L. proposed and designed the research. G.C. contributed in sample preparation. Y.L., L.Y., X.W.J., Y.Y.P., C.Y.C., Z.J.X., D.X.M., J.F.H., X.L., Y.F., H.M.Y., L.Z., G.D.L., S.L.H., X.L.D., J.Z., Z.Y.X., C.T.C. and X.J.Z. contributed to the development and maintenance of Laser-ARPES system. Y.L. carried out the ARPES experiment.  Y.F., L.Y. and X.J.Z. analyzed the data. J.Z.Z., H.M.W., X.D. and Z.F. performed band structure calculations.  X.J.Z., Y.L. and L.Y. wrote the paper and all authors participated in discussion and comment on the paper.

\vspace{3mm}

\noindent {\bf\large Additional information}\\

\noindent{\bf Competing financial interests:} The authors declare no competing financial interests.

\newpage

\begin{figure*}[tbp]
\begin{center}
\includegraphics[width=1.0\columnwidth,angle=0]{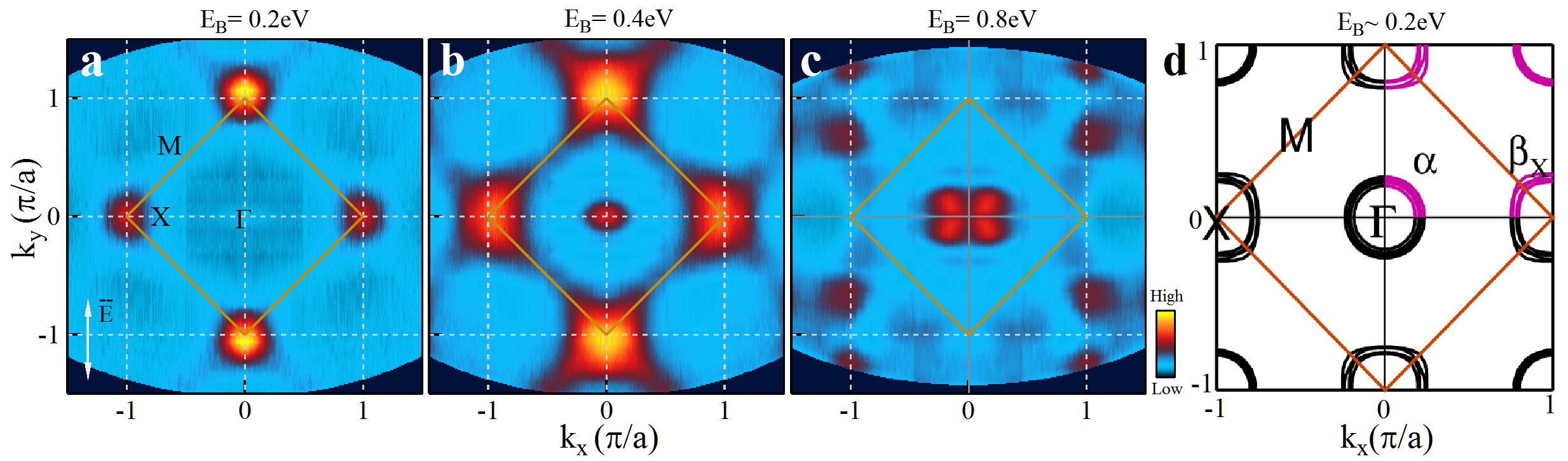}
\end{center}
\caption{{\bf Measured constant energy contours of Sr$_2$IrO$_4$ and its comparison with calculations.} (a-c) represent constant energy contours of the spectral weight distribution for  Sr$_2$IrO$_4$ measured at $\sim$20 K at different binding energies (E$_B$) of 0.2 eV, 0.4 eV, and 0.8eV, respectively. (d) is the calculated constant energy contour at a binding energy of $\sim$0.2 eV by including on-site Coulomb repulsion and spin-orbit coupling\cite{BJKim2008PRL}. The orange lines denote the antiferromagnetic Brillouin zone boundary for the IrO$_2$ plane.
}
\end{figure*}

\newpage

\begin{figure*}[tbp]
\begin{center}
\includegraphics[width=1.0\columnwidth,angle=0]{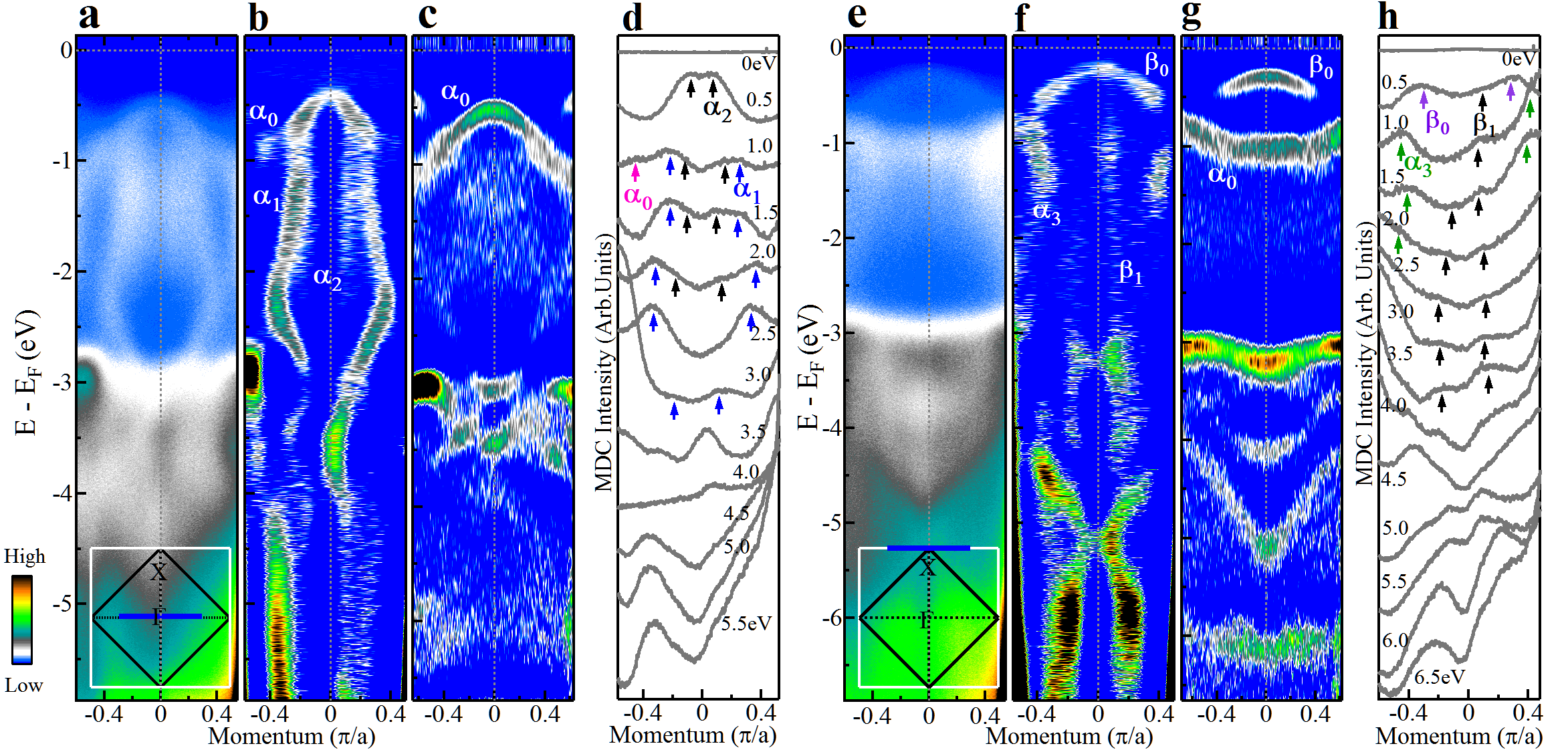}
\end{center}
\caption{{\bf Typical band structures of Sr$_2$IrO$_4$ along high-symmetry cuts in a large energy range.} (a) Original photoemission image of Sr$_2$IrO$_4$ measured along a high-symmetry cut across $\Gamma$; the location of the cut is shown as a solid blue line in the inset. (b) and (c) are corresponding momentum-second-derivative and energy-second-derivative images of (a), respectively. (d) Momentum distribution curves (MDCs) at different binding energies obtained from (a). (e). Original photoemission image measured along a high-symmetry cut across X; the location of the cut is shown as a solid blue line in the inset. (f) and (g) are corresponding momentum-second-derivative and energy-second-derivative images of (e), respectively. (h) MDCs at different binding energies obtained from (e).
}
\end{figure*}

\newpage

\begin{figure*}[tbp]
\includegraphics[width=1.0\columnwidth,angle=0]{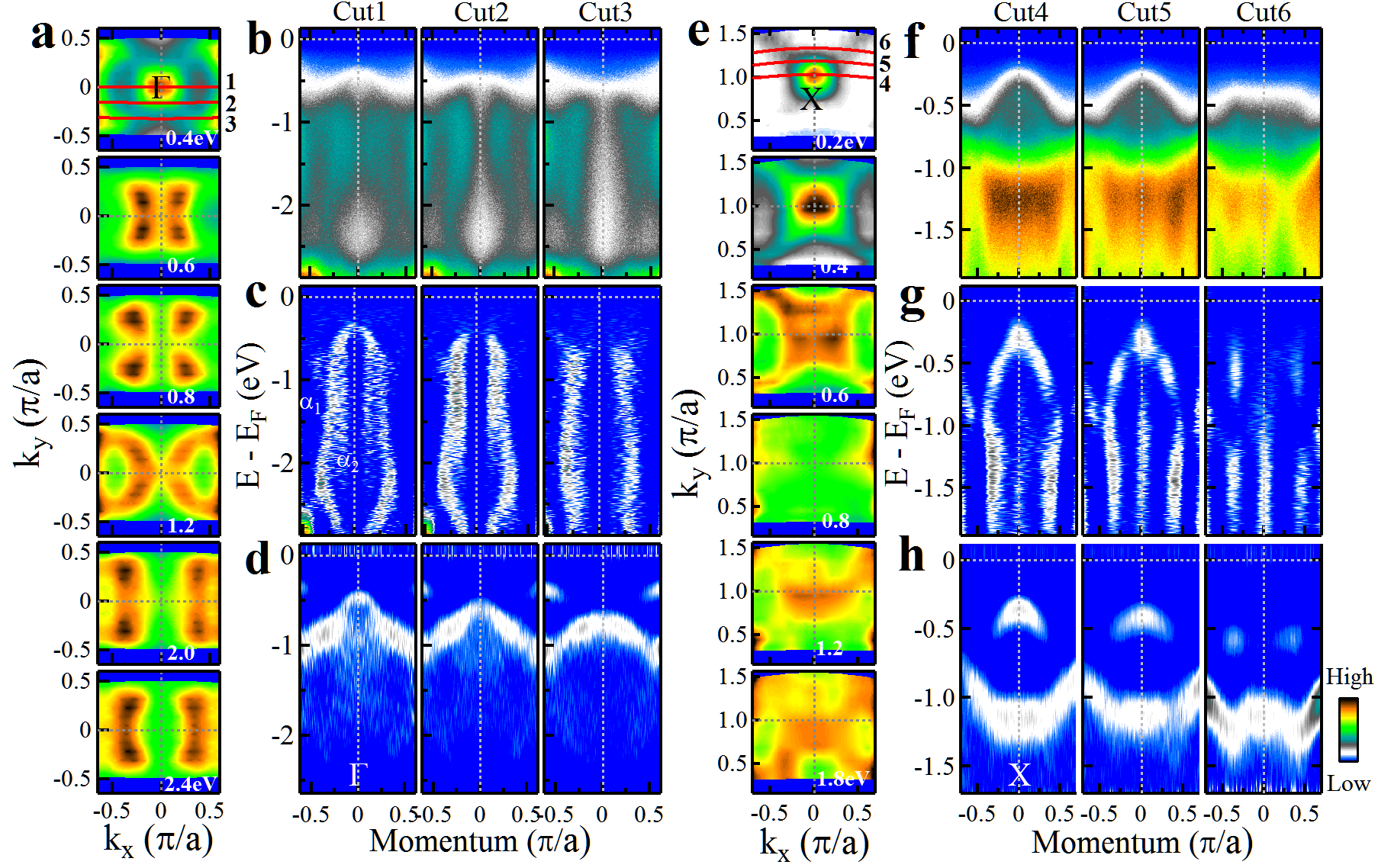}
\begin{center}
\caption{{\bf Momentum dependence of the band structures around $\Gamma$ and X regions.} (a) Constant energy contours around $\Gamma$ point at different binding energies from 0.4 eV (top panel)£¬ to 0.6, 0.8, 1.2, 2.0 and 2.4 eV (bottom panel). (b) Original photoemission images measured along different momentum cuts around $\Gamma$. The location of the momentum cuts are shown as red lines in the top panel of (a).  (c) and (d) are corresponding momentum-second-derivative and energy-second-derivative images of (b), respectively. (e) Constant energy contours around X point at different binding energies from 0.2 eV (top panel) to 0.4, 0.6, 0.8, 1.2 and  1.8 eV (bottom panel). (f) Original photoemission images measured along different momentum cuts around X. The location of the momentum cuts are shown as red lines in the top panel of (e).  (g) and (h) are corresponding momentum-second-derivative and energy-second-derivative images of (f), respectively.
}
\end{center}
\end{figure*}

\newpage

\begin{figure*}[tbp]
\begin{center}
\includegraphics[width=1.0\columnwidth,angle=0]{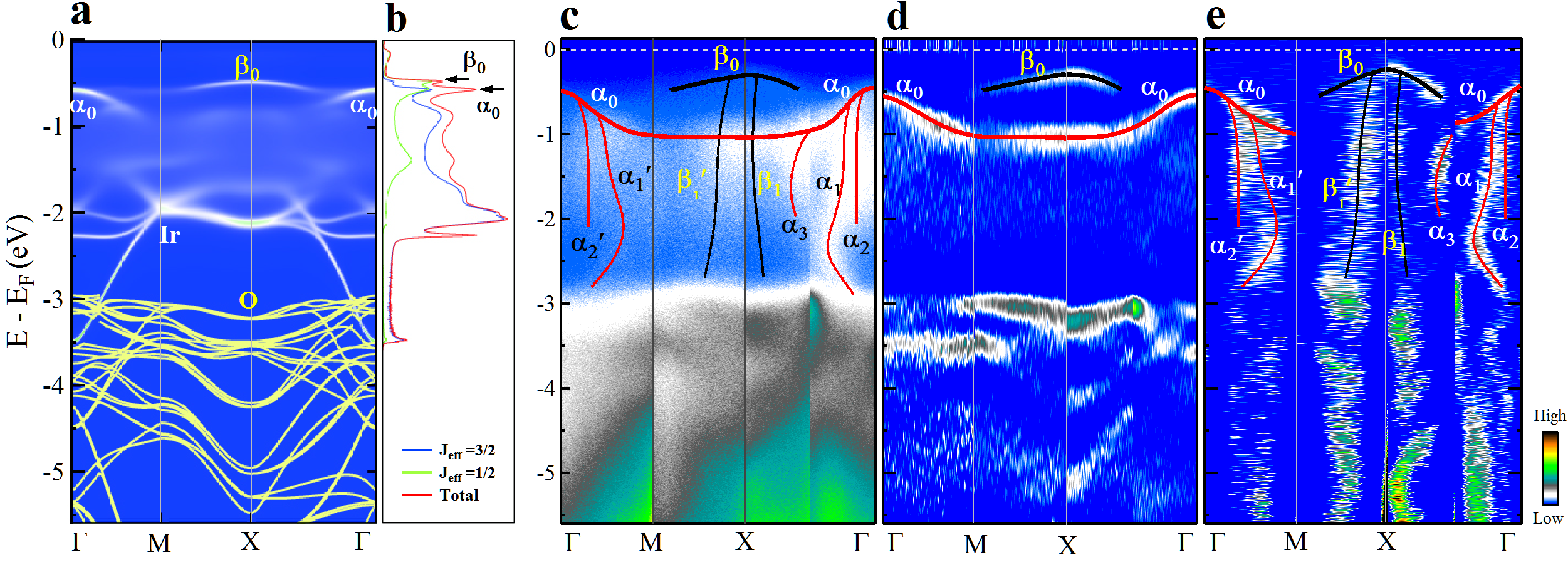}
\end{center}
\caption{{\bf Calculated and measured overall band structure of Sr$_2$IrO$_4$.} (a) Band structure of Sr$_2$IrO$_4$ by DMFT calculations along high symmetry line in the first Brillouin zone. The white lines are the LDA+DMFT calculation on Iridium's t$_{2g}$ orbitals while the yellow lines are the LDA calculation on Oxygen p orbitals.  (b) Calculated density-of-states for the J$_{eff}$=3/2 and J$_{eff}$=1/2 states, and the total density-of-states of the Iridium orbitals. (c). Overall measured original photoemission image of Sr$_2$IrO$_4$ along high-symmetry cuts. The observed bands are overlaid on top of the original data. (d) and (e) are corresponding energy-second-derivative and momentum-second-derivative images of (c), respectively. The black and red lines are guides to the eye for the bands that can be resolved.  $\alpha_1^{'}$, $\alpha_2^{'}$ and $\beta_1^{'}$ bands are equivalent bands to the $\alpha_1$, $\alpha_2$, and $\beta_1$ bands along other symmetry cuts.
}
\end{figure*}

\end{document}